\documentclass[Journal]{IEEEtran}
\usepackage{amsmath}
\usepackage{amssymb}
\usepackage{amsfonts}
\usepackage[final]{graphicx}
\usepackage{ctable}
\usepackage{algorithm}
\usepackage{algorithmic}
\usepackage{multirow}
 \usepackage{subfigure}
 \usepackage{url}

 \title{ Energy-Efficient  UAV  Multicasting  with Simultaneous FSO Backhaul  and  Power Transfer }
 \author{Yue Ling Che,~\IEEEmembership{Member,~IEEE,} Weibin Long,  Sheng Luo, \IEEEmembership{Member,~IEEE,}
 Kaishun Wu,~\IEEEmembership{Member,~IEEE,} \\and Rui Zhang,~\IEEEmembership{Fellow,~IEEE} \vspace{-0.15in}
 \thanks{Y. L. Che, W. Long, S. Luo, and K. Wu are with the  College of Computer Science and Software Engineering, Shenzhen University, China (e-mail: yuelingche@szu.edu.cn, 1800271002@email.szu.edu.cn, \{sluo,wu\}@szu.edu.cn).}
 \thanks{R.~Zhang is with the Department of Electrical and Computer Engineering,
 National University of Singapore (e-mail: elezhang@nus.edu.sg). }
 }

\begin{document}

\maketitle
\thispagestyle{empty}

\begin{abstract}

 This letter studies an  unmanned aerial vehicle (UAV)   aided multicasting (MC) system, which is  enabled by simultaneous free space optics (FSO) backhaul and power transfer. The UAV applies the power-splitting  technique to harvest wireless power and decode backhaul information simultaneously over the FSO link, while at the same time using the harvested   power to multicast the backhauled information over the radio frequency (RF) links to  multiple   ground users (GUs). We derive the UAV's achievable MC rate under the Poisson point process (PPP) based GU distribution.  By jointly designing the FSO and RF links and the UAV altitude, we maximize the system-level energy efficiency (EE), which  can be equivalently expressed as the ratio of the UAV's MC rate over the optics base station (OBS) transmit power,  subject to the UAV's sustainable operation and reliable backhauling constraints. Due to the non-convexity of this problem, we propose suboptimal solutions with low complexity. Numerical results show the close-to-optimal EE performance by properly balancing the power-rate tradeoff between    the   FSO power and  the MC data transmissions.

\end{abstract}


\newtheorem{definition}{\underline{Definition}}[section]
\newtheorem{fact}{Fact}
\newtheorem{assumption}{Assumption}
\newtheorem{theorem}{\underline{Theorem}}[section]
\newtheorem{lemma}{\underline{Lemma}}[section]
\newtheorem{corollary}{\underline{Corollary}}[section]
\newtheorem{proposition}{\underline{Proposition}}[section]
\newtheorem{example}{\underline{Example}}[section]
\newtheorem{remark}{\underline{Remark}}[section]
\newcommand{\mv}[1]{\mbox{\boldmath{$ #1 $}}}
\newtheorem{property}{\underline{Property}}[section]

\section{Introduction}

The rapid growth of unmanned aerial vehicles (UAVs)  has  brought  fertile applications in  wireless communications  \cite{Zeng.IEEE.2019}.
Although appealing, the feasibility of the UAV-aided wireless communications still faces crucial  challenges. First,  due to the   limited battery energy on-board,  the UAV's mission time is severely constrained.  The technique  of solar power harvesting   can be  adopted to prolong the UAV's battery lifetime, but the intermittent solar energy availability is a practical issue \cite{Kuan.TCOM}.  One the other hand,  it is also  difficult to establish reliable  wireless backhaul links for the  UAVs. Due to the presence of strong air-to-ground (A2G) line-of-sight (LoS) links,  the  conventional radio frequency (RF) based wireless backhaul transmission  from the terrestrial base station  (BS) to the UAV  may suffer  severe   interference from the co-channel terrestrial transmissions, and vice versa \cite{Qiu.TCOM.2020}.

 \begin{figure}
\centering
\DeclareGraphicsExtensions{.eps,.mps,.pdf,.jpg,.png}
\DeclareGraphicsRule{*}{eps}{*}{}
\includegraphics[angle=0,width=0.48\textwidth]{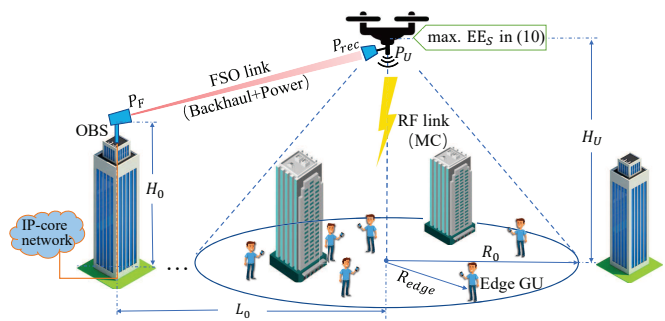}
\vspace{-0.1in}
\caption{UAV-aided MC   with   simultaneous FSO backhaul and power transfer.}
\label{fig: system_model}
\vspace{-0.2in}
\end{figure}

The  technology of free space optics (FSO)  transmission is promising    to address the above two challenges.
Due to the  narrow  beam with intense energy concentration, the FSO beam is   able to   deliver high power (of, e.g.,  hundreds of Watts) to the UAV   over a long  range \cite{Ouyang.ICC.2018}, \cite{Zhao.TWC.2020}.
By exploiting the large terahertz  unlicensed bandwidth, the FSO   link also achieves high-speed  data rates, without interfering the UAV's RF communication \cite{Schulz.LighWJ.2016}-\cite{Wang.TWC.2020}.
Moreover, the simultaneous FSO information and power transfer has been investigated in  \cite{Wang.JSAC.2015}  and  \cite{Lahmeri.Cletter.2020}.
In addition,  with    the emerging FSO systems developed with fast tracking, the  UAV performance loss due to  FSO pointing errors can be effectively compensated  \cite{FSO_book}.

In this letter,   we  propose  a novel UAV-aided downlink multicasting (MC)  system with simultaneous FSO backhaul  and power transfer, as shown in Fig. 1.  Adopting the power-splitting based FSO receiver \cite{Wang.JSAC.2015},  the UAV   harvests  FSO power and decodes  backhaul information simultaneously in the FSO link, while at the same time using the harvested FSO power to  multicast the backhauled information to multiple ground users (GUs) in the RF  links.
Under the homogeneous Poisson point process (PPP) based GU distribution, we derive the UAV's  achievable MC rate and the mission  completion time using the practical LoS probability-based A2G  channel model   \cite{Hourani.letter.2014}.

To assure the UAV's  sustainable and reliable MC over the RF link, an  optics base station (OBS) that transmits power  and data for charging and backhauling the UAV over the FSO link is essential.
We  thus aim to maximize the  OBS's energy efficiency (EE) at a system level, which is defined as the ratio of the UAV's  MC rate over the OBS transmit power,
by jointly optimizing the OBS  transmit power  and the   power-splitting ratio for the FSO link, the UAV transmit power for the RF link, and the UAV  altitude,  subject to  the UAV's   FSO power harvesting (FPH)  and   FSO backhauling rate (FBR) constraints.
This optimization problem is   non-convex, and thus difficult to be solved optimally.
  Despite of that, we obtain the optimal OBS  transmit power and  power-splitting ratio   in closed-forms   under arbitrarily given UAV transmit power and altitude.   Then, by proposing a tight approximation   to the system  EE,  we  obtain suboptimal UAV transmit power and altitude with low complexity.  Numerical results   show the close-to-optimal  EE performance, which  achieves a proper power-rate tradeoff between maximizing  the   FSO power and  the MC data rate.

 In the literature,  the   UAV communications have been studied in  \cite{Ouyang.ICC.2018} and \cite{Zhao.TWC.2020}  with FSO power transfer, or \cite{Schulz.LighWJ.2016} and \cite{Ajam.Open} with FSO backhauling, respectively. It is noted that     only limited  works  such as  \cite{Lahmeri.Cletter.2020} and  \cite{GlobeCom} have considered simultaneous FSO backhaul and power transfer enabled UAV communication systems,  while they do not address how to efficiently use the UAV's harvested FSO power for optimizing its  RF communication performance.
 To our best knowledge, our proposed    energy-efficient UAV MC by jointly designing   the UAV's FSO and RF links under
 practical UAV power consumption  and A2G channel models  is  new and  has not been reported in the literature yet.

\section{System Model}

As shown in Fig. 1,  a set of GUs are distributed   following a homogeneous  PPP $\Psi(\lambda_G)$ of density $\lambda_G>0$  within the area of $\phi(\mv o, R_0)$, where $\phi(\mv o, R_0)$ represents a circular area of radius $R_0$  centered at the origin $\mv o\!=\!(0,0)$ on the horizontal plane. We say   GU-$i$ locating at $\mv x_i \! \in \! \Psi(\lambda_G)$   is of radius $r$ if $\| \mv x_i\|\!=\!r$, where  $\|\! \cdot\! \|$ is the  Euclidean norm.
The UAV hovers   above  $\mv o$  for achieving high-quality RF communication, and is installed with  a power-splitting based FSO receiver  and an RF transmitter.
To avoid   transmission blockage, the OBS is  deployed at a high altitude (e.g., on top of a building \cite{Schulz.LighWJ.2016}) of   $H_0$.
Denote the UAV's altitude as $H_U\! \in \! [H_0, H_{\textrm{max}}]$, and its horizontal   distance to the OBS  as $L_0 \!\geq  \!0$.

\subsection{Downlink RF Multicasting}
Under  the LoS-probability based A2G channel  in \cite{Hourani.letter.2014} and \cite{Mamaghani.Access.2019}, we use  a binary indicator with $\mu(r)\!=\!1$ or $\mu(r)\!=\!0$ to  denote  the occurrence of LoS or NLoS link from the UAV to a GU at radius $r$,   respectively. The A2G channel      $h(r)$    is thus given by
 \begin{small}
 \begin{equation} \label{eq: A2G_channel}
 h(r)= \begin{cases}
\left(H_U^2+r^2\right)^{-\frac{\alpha_{\textrm{L}}}{2}},  \textrm{~if~}\mu(r)=1,  \\
\left(H_U^2+r^2\right)^{-\frac{\alpha_{\textrm{N}}}{2}},  \textrm{~if~}\mu(r)=0,
 \end{cases}
 \end{equation}
 \end{small}where  $\alpha_{\textrm{L}}$ and $\alpha_{\textrm{N}}$ with $\alpha_{\textrm{L}}\! < \!\alpha_{\textrm{N}}$ denote  the LoS and NLoS channel path-loss exponents, respectively.
We adopt  the   LoS probability $\mathcal{P}_{\textrm{L}}(r)\!\triangleq \!\mathbb{P}(\mu(r)\!=\!1) \!=\!\frac{1}{1+a\exp(-b(\theta(r)-a))}$,
where $a$ and $b$ are environment-related constant parameters, and $\theta(r)\! =\! \frac{180}{\pi}\arctan(\frac{H_U}{r})$ is the A2G elevation angle.
 The NLoS probability is thus obtained as $\mathcal{P}_{\textrm{N}}(r)=1\!-\!\mathcal{P}_{\textrm{L}}(r)$.

 Let $P_U\!\in \! [0, P_{\textrm{max}}]$  denote the UAV's transmit power, where $P_{\textrm{max}}$ is the UAV's maximally allowable transmit power. The GU's received signal-to-noise ratio (SNR) over the LoS or NLoS link  is given as  $\Gamma_{\textrm{L}}(r)\!=\!\frac{P_U}{\sigma_{\textrm{L}}^2} (H_U^2 \!+\! r^2)^{\!-\!\frac{\alpha_{\textrm{L}}}{2}} $ or $\Gamma_{\textrm{N}}(r)\!=\!\frac{P_U}{\sigma_{\textrm{N}}^2} (H_U^2  +   r^2)^{\!- \frac{\alpha_{\textrm{N}}}{2}} $, respectively, where $\sigma_{\textrm{L}}^2$ and $\sigma_{\textrm{N}}^2$ are the GU's received noise power strength in LoS and NLoS link, respectively. Denoting the bandwidth of the RF link as $B$,  the
 average   transmission rate (in bits/s)  achieved at a GU  at radius $r\in[0,R_0]$  is obtained as
 \begin{small}
 \begin{equation}\label{eq: C}
\bar{C}(r)= \mathcal{P}_{\textrm{L}}(r) B \log_2\!\left(1\!+\!\Gamma_{\textrm{L}}(r)  \right)\!+\!  \mathcal{P}_{\textrm{N}}(r) B \log_2\!\left( 1\!+\!\Gamma_{\textrm{N}}(r) \right).
\end{equation}
 \end{small}It is easy to show that a lower bound  of $\bar{C}(r)$ is given by
 \begin{small}
 \begin{equation}\label{eq: C_low}
\bar{C}_{low}(r)\!\triangleq \!\mathcal{P}_{\textrm{L}}(r) B \log_2\!\left(1+\Gamma_{\textrm{L}}(r)  \right) \leq \bar{C}(r).
\end{equation}
 \end{small}As will be shown  in Section IV by simulations,  for any given $R_0>0$ and $r\in[0,R_0]$,   since $\mathcal{P}_{\textrm{L}}(r)$ increases (and thus $\mathcal{P}_{\textrm{N}}(r)$ decreases) over $H_U$, which leads to $\mathcal{P}_{\textrm{L}}(r)\!>\!\mathcal{P}_{\textrm{N}}(r)$ when $H_U$ is large,
 and  also due to  $\alpha_{\textrm{L}}\! < \!\alpha_{\textrm{N}}$ and  thus $\Gamma_{\textrm{L}}\! > \!\Gamma_{\textrm{N}}$ in general,  $\bar{C}_{low}(r)$ is generally close to $\bar{C}(r)$,   when $H_U$  is sufficiently large so that  $H_U\geq H_0$. We hence use $\bar{C}_{low}(r)$ to approximate the more complicated $\bar{C}(r)$ in the following.

 Moreover, suppose that the UAV's  common information file for downlink MC is  of    $\bar{D}$ bits.
 As $\bar{C}_{low}(r)$ decreases over $r$,  the UAV's downlink MC  is completed if the \emph{edge GU} at radius $R_{edge}= \max_{\mv x_i \in \Psi(\lambda_G)}   \| \mv x_i\|  $ receives the complete file of  $\bar{D}$ bits.
Based on the null-probability of the PPP, the cumulative distribution function (CDF) of $R_{edge}$ is obtained as \cite{Che.JSAC.2015}
 \begin{small}
 \begin{align} \label{eq: CCF_R_edge}
 \mathbb{P}(R_{edge}\leq r) &= \mathbb{P}(\textrm{no~GUs~exist~within~} \phi(\mv o, R_0)-\phi(\mv o, r)) \nonumber \\
 &=e^{-\pi \lambda_G(R_0^2-r^2)},
  \end{align}
  \end{small}where the circular ring $ \phi(\mathbf{o}, R_0)\!-\!\phi(\mathbf{o}, r)$ is of radius between  $r$ and $R_0$. By differentiating
  (\ref{eq: CCF_R_edge}) over $r$, the probability density function (pdf) of $R_{edge}$ is obtained as $ f_{R_{edge}}(r)=e^{-\pi \lambda_G(R_0^2-r^2)} 2\pi \lambda_G r$.
  The average downlink transmission rate  achieved  at the edge GU is hence given by
 \begin{small}
  \begin{align} \label{eq: C_edge}
 &\bar{C}_{edge}= \int_{r=0}^{R_0}  \bar{C}_{low}(r)  f_{R_{edge}}(r)\, dr \nonumber \\
 &= \int_{r=0}^{R_0} \! \mathcal{P}_{\textrm{L}}(r) B \log_2\!\left(1\!+\!\Gamma_{\textrm{L}}(r)  \right)\! e^{-\pi \lambda_G(R_0^2-r^2)} 2\pi \lambda_G r \, dr.
  \end{align}
  \end{small}As a result, the average mission completion time for the UAV's downlink MC can be approximated as $T_{MC}=\bar{D}/\bar{C}_{edge}$.

\subsection{Simultaneous  FSO  Backhaul and Power Transfer}
We now consider the FSO  transmissions, where the intensity modulation/direct detection (IM/DD) is applied \cite{Schulz.LighWJ.2016}.
The OBS-to-UAV distance is obtained as  $L_{back}\!=\!\sqrt{L_0^2\!+\!(H_U\!-\!H_0)^2}$. Let      $D_{\textrm{r}}$  and  $\theta_{\textrm{t}}$ denote the  diameter of the FSO receiver at the UAV   and  the full transmitting divergence angle, respectively.
From  \cite{Najafi.TCOM.2020}, the FSO  channel path loss is mainly affected by the following three phenomena: 1) the FSO receiver's responsivity, which is represented by a constant $ \tau_{\textrm{e}}\in(0,1)$; 2) the atmospheric turbulence induced fading, which is modeled as   $\frac{D_{\textrm{r}}^2}{\theta_{\textrm{t}}^2 L_{back}^2} 10^{\!-\kappa \frac{L_{back}}{10}} $ with a    weather-dependent coefficient $\kappa $; and 3) the random geometric and misalignment loss  (GML), which is caused by the UAV's  fluctuations  around its hovering location.
In this letter, we focus on the  average received power strength at the UAV, and  use $\tau_{\textrm{GML}}$ to represent the average  GML for the FSO channel, which can be obtained numerically based on the GML's pdf   derived in \cite{Najafi.TCOM.2020}.
As a result,  denoting the OBS transmit power   as $P_F$,  the average received   FSO power   at the UAV   is expressed as
\begin{small}
\begin{equation}\label{eq: P_rec}
P_{rec}=P_F \omega (H_U)=P_F\tau_{\textrm{e}} \tau_{\textrm{GML}} \frac{D_{\textrm{r}}^2}{\theta_{\textrm{t}}^2 L_{back}^2} 10^{-\kappa L_{back}/10},
\end{equation}
\end{small}where   $\omega (H_U)\!= \! \frac{\tau_{\textrm{e}} \tau_{\textrm{GML}} D_{\textrm{r}}^2}{\theta_{\textrm{t}}^2 L_{back}^2} 10^{\!-\kappa \frac{L_{back}}{10}}$ is the  average FSO channel  power gain.

Next, let  $\rho\!\in\![0,1]$ denote the power-splitting ratio, where  $P_{rec} \rho$  amount of power is used to decode backhaul information,  and  the remaining power of $P_{rec}(1-\rho)$   is for UAV power harvesting.  For the IM/DD FSO   channel, the theoretical information capacity remains unknown \cite{Lapidoth.IIT.2009}. However, as  validated by the test fields in  \cite{Schulz.LighWJ.2016}, the  achievable FSO  throughput  at the UAV can be properly expressed  as
\begin{small}
\begin{equation}\label{eq: D_back}
D_{back}=T_{MC} \frac{W}{2} \log_2 \left(1+\frac{P_F \omega (H_U) \rho}{\sigma_U^2 \beta}   \right),
\end{equation}
\end{small}where  $T_{MC}$ is applied since  the UAV operates over the FSO and  RF links at the same time,  $W\!>\!B$ is the FSO channel bandwidth, $\sigma_U^2$ is the UAV's received  noise power,  and the division by $2$ is  due to  the  real-valued Gaussian channel  for the IM/DD FSO link. Particularly,  the  empirical factor  $\beta$ in (\ref{eq: D_back}) is introduced to include all the  implementation-related   SNR losses in practice, which includes,  e.g., the reduced modulation current for the IM/DD channel, the  non-ideal channel coding,  and the   FSO pointing errors.  It is also noted  that when $\beta\!=\!\frac{2\pi}{e}$, $D_{back}$ in (\ref{eq: D_back}) becomes the well-known  FSO throughput lower-bound in    \cite{Lapidoth.IIT.2009}, without considering the above mentioned implementation-related   SNR losses. We hence assume that $\beta\!\geq \!\frac{2\pi}{e}$ in this letter. This is  also in accordance with the  empirical value $\beta\!=\!15$ dB  obtained  in   \cite{Schulz.LighWJ.2016}.  Since $D_{back}\!\geq \!\bar{D}\!=\!T_{MC}\bar{C}_{edge}$ is required for reliably backhauling the UAV,   the FBR   constraint is  given as
\begin{small}
\begin{equation} \label{eq: backhaul_constraint}
\frac{W}{2} \log_2 \left(1+\frac{P_F \omega (H_U) \rho}{\sigma_U^2 \beta}\right) \geq \bar{C}_{edge}.
\end{equation}
\end{small}
$~~$Moreover,  we consider a linear FPH model based on \cite{Ouyang.ICC.2018} and \cite{Zhao.TWC.2020},  where the UAV's  harvested FSO power is  expressed as $\eta P_{rec}  (1\!-\!\rho)$ with constant FPH efficiency $\eta \!\in \! (0,1)$. The complicated non-linear FPH efficiency will be studied in our future work. Let $P_{\textrm{hov}}$ denote the UAV's propulsion power  for hovering  with $P_{\textrm{hov}}\!\gg \!P_U$ in practice from \cite{Che.TWC.2021} and \cite{Yang.TVT.2020}. To assure the UAV's sustainable operation, we obtain
the UAV's FPH constraint as follows:
\begin{small}
\begin{equation}\label{eq: LEH_constraint}
\eta P_F \omega (H_U) (1-\rho)   \geq P_{\textrm{hov}}   +P_U.
\end{equation}
\end{small}$~~~$At last,   define the system-level  energy efficiency    as $\textrm{EE}_S\!\triangleq \!\bar{D}/E_{total}$, where
$E_{total}=P_F T_{MC}$ is the overall FSO energy consumption for backhauling and charging the UAV over the FSO link, such that the UAV's  MC mission over the RF link to deliver $\bar{D}$ bits of common file can be completed within $T_{MC}${\footnote{ To focus on the joint design of the FSO and RF links, the OBS's and  GUs's circuit energy consumptions are not considered.}}.
Since  the common information file is completely delivered to all GUs within the corresponding time of   $T_{MC}\!=\!\bar{D}/\bar{C}_{edge}$, regardless of its size $\bar{D}$, $\textrm{EE}_S$ can be further expressed as
\begin{small}
\begin{equation}\label{eq: EE}
\textrm{EE}_S\!=\!\frac{ \bar{C}_{edge} }{P_F}.
\end{equation}
\end{small}

\section{ System-Level EE Maximization}
In this  section, we   aim to maximize $\textrm{EE}_S$ in (\ref{eq: EE}), by jointly optimizing the OBS transmit power $P_F$, the power-splitting ratio $\rho$, the UAV's altitude $H_U$ and transmit power $P_U$, subject to the FBR constraint in (\ref{eq: backhaul_constraint}) and the FPH constraint in (\ref{eq: LEH_constraint}) at the UAV. The optimization problem is  formulated as follows.
\begin{small}
\begin{align}
{\textrm{(P1)}}:~\max_{P_F, \rho, \atop H_U, P_U} ~& \frac{ \bar{C}_{edge} }{P_F} \nonumber \\
\mathrm{s.t.}~& 0\leq \rho\leq 1, 0\leq P_U \leq P_{\textrm{max}}, P_F>0,  \nonumber \\
& H_0\leq H_U \leq H_{\textrm{max}}, (\ref{eq: backhaul_constraint}), (\ref{eq: LEH_constraint}). \nonumber
\end{align}
\end{small}Due to the complicated expression of $\bar{C}_{edge}$  in (\ref{eq: C_edge}) and coupled decision variables,  problem (P1) is non-convex in general. To solve problem (P1) efficiently, in the following,   we first obtain the optimal $P_F$ and $\rho$  under any given  $H_U$ and $P_U$, and then derive  the optimal $H_U$ and $P_U$.

First, we  optimize  $P_F$ and $\rho$ for the FSO link under any given $H_U$ and $P_U$. In this case,  both of the UAV's power consumption $P_{U}+P_{\textrm{hov}}$ and the MC rate $ \bar{C}_{edge}$ over the RF link become fixed. As a result, maximizing $ \frac{ \bar{C}_{edge} }{P_F} $ in problem (P1) is equivalent to minimizing   $P_F$  over the FSO link  subject to (\ref{eq: backhaul_constraint}) and (\ref{eq: LEH_constraint}), which leads to the following proposition.

\begin{proposition} \label{proposition: optimal_FSO}
Let $Q_I\!=\!\frac{\sigma_U^2 \beta}{\omega(H_U)} \big(2^{\frac{2}{W}\bar{C}_{edge}}\!-\!1  \big)$ and $Q_E\!=\!\frac{P_U + P_{\textrm{hov}}}{\eta\omega(H_U)}$. For  any arbitrarily given $P_U\in [0, P_{\textrm{max}}]$ and $H_U\in [H_0, H_{\textrm{max}}]$  at the UAV,  the optimal $P_F(P_U,H_U)$ and $\rho(P_U,H_U)$ for the FSO link are obtained in closed-forms as
\begin{small}
\begin{flalign}\label{eq: optimal_FSO}
P_F(P_U,H_U)=Q_I+Q_E,~~ \rho(P_U,H_U)=\frac{Q_I}{Q_I+Q_E}.
\end{flalign}
\end{small}
\end{proposition}

Proposition \ref{proposition: optimal_FSO} can be shown by  noting from  (\ref{eq: backhaul_constraint}) and (\ref{eq: LEH_constraint})  that $Q_I\leq P_F \rho$ and $Q_E\leq P_F (1-\rho)$, respectively, and thus   $P_F \geq Q_I+Q_E$, which yields    (\ref{eq: optimal_FSO}) to  maximize $ \frac{ \bar{C}_{edge} }{P_F}$ under a fixed $\bar{C}_{edge} $. Its proof is thus omitted for brevity.

From (\ref{eq: optimal_FSO}),    both  constraints in (\ref{eq: backhaul_constraint}) and   (\ref{eq: LEH_constraint})  hold with equalities under the optimal   $P_F(P_U,H_U)$  and   $\rho(P_U,H_U)$. Hence,  all the UAV's harvested FSO power  is consumed  completely  to support its hovering and simultaneous communication over the RF links for EE maximization.

Next, we consider the  RF link.
Under the optimal  $P_F(P_U,H_U)$ and $\rho(P_U,H_U)$ in Proposition \ref{proposition: optimal_FSO}, problem (P1) is reduced to  the following problem:
\begin{small}
\begin{align}
{\textrm{(P2)}}:\!\max_{P_U,H_U}&  \textrm{EE}_U\!\triangleq\!  \frac{ \bar{C}_{edge} }{\frac{\sigma_U^2 \beta}{ \omega(H_U)} (2^{\frac{2}{W}\bar{C}_{edge}}\!\!-\!1)\!+\!\!\frac{P_U\!+\!P_{\textrm{hov}}}{\eta  \omega(H_U)}}   \label{eq: EE_U} \\
\mathrm{s.t.}~&  0\leq P_U \leq P_{\textrm{max}}, H_0\leq H_U \leq H_{\textrm{max}},  \nonumber
\end{align}
\end{small}where  $\textrm{EE}_U$ is obtained by substituting (\ref{eq: optimal_FSO})  into (\ref{eq: EE}). (P2) is  non-convex  due to  the fractional form of the objective with complicated $\bar{C}_{edge}$.
To deal with the complicated $\bar{C}_{edge}$, an upper-bound of  $\bar{C}_{edge}$ is   proposed as follows.
\begin{proposition}\label{proposition: C_upper_bound}
Define $\bar{C}_{edge}^{upp}\triangleq  \mu B\log_2\left(1+P_U y(H_U)\right) $,  where $\mu= \int_{r=0}^{R_0} \! \mathcal{P}_{\textrm{L}}(r)  e^{-\pi \lambda_G(R_0^2-r^2)} 2\pi \lambda_G r \, dr$ and $y(H_U)= \frac{1}{\sigma_{\textrm{L}}^2}\int_{r=0}^{R_0} \! \frac{1}{\mu}\mathcal{P}_{\textrm{L}}(r)  (H_U^2+r^2)^{-\frac{\alpha_{\textrm{L}}}{2}} e^{-\pi \lambda_G(R_0^2-r^2)} 2\pi \lambda_G r \, dr$.
When $H_U>\sqrt{\alpha_{\textrm{L}}+1}R_0$, $\bar{C}_{edge} \leq \bar{C}_{edge}^{upp}$ holds.
\end{proposition}
\begin{IEEEproof}
Please refer to Appendix A.
\end{IEEEproof}

By replacing $\bar{C}_{edge} $  in (\ref{eq: EE_U}) with $\bar{C}_{edge}^{upp}$, problem (P2) is transformed into the following problem.
\begin{small}
\begin{align}
\! {\textrm{(P3)}}: \! \max_{P_U,H_U}  &\widetilde{\textrm{EE}}_U\!\triangleq\!\frac{ \bar{C}_{edge}^{upp} }{\frac{\sigma_U^2 \beta}{ \omega(H_U)} (2^{\frac{2}{W}\bar{C}_{edge}^{upp}}\!\!-\!1)\!+\!\!\frac{P_U\!+\!P_{\textrm{hov}}}{\eta  \omega(H_U)}}  \label{eq: EE_U_approx}\\
&~~~~~=\!  \frac{ (\sigma_U^2 \beta)^{-1}\mu B \omega(H_U) \log_2\left(1\! +\! P_U y(H_U)\right) }{ (1\! +\! P_U y(H_U)^{\!\frac{2\mu B}{W}}\! \!+\! \frac{P_U\! +\! P_{\textrm{hov}}}{\eta\sigma_U^2 \beta}\! -\! 1}  \label{eq: EE_U_approx_complete} \\
\mathrm{s.t.}~&  0\leq P_U \leq P_{\textrm{max}}, H_0\leq H_U \leq H_{\textrm{max}}.  \nonumber
\end{align}
\end{small}
It is observed that $\textrm{EE}_U$  in  (\ref{eq: EE_U})  and  $\widetilde{\textrm{EE}}_U$  in  (\ref{eq: EE_U_approx})  are of the same fractional structure with a  practically large $P_{\textrm{hov}}$     in the denominator, while
 the numerator   and    denominator  both increase  over  $\bar{C}_{edge}$ or $\bar{C}_{edge}^{upp}$.
  It is thus easy  to verify that both $\textrm{EE}_U$  and  $\widetilde{\textrm{EE}}_U$ are  more dominated by the   large $P_{\textrm{hov}}$   than the corresponding edge GU  rates.
 As a result,
  although $\bar{C}_{edge}\!\leq \!\bar{C}_{edge}^{upp}$ holds under a large $H_U$, as will be shown in Section IV by simulations, $\widetilde{\textrm{EE}}_U$    is generally a tight  approximation to  $\textrm{EE}_U$  regardless of  $H_U$'s values,  under  the optimal $P_U$ to problem (P3) and (P2), respectively.

From (\ref{eq: EE_U_approx_complete}) under $\bar{C}_{edge}^{upp}$'s expression,  although $\widetilde{\textrm{EE}}_U$   is  neither convex nor concave over $(P_U,H_U)$,  the optimal $P_U(H_U)$ at any given $H_U$ to problem (P3)  is obtained as  follows.
 \begin{proposition}\label{proposition: suboptimal_P_U}
 For  any given $H_U$, let $g(P_U)\triangleq\frac{\partial \widetilde{\textrm{EE}}_U}{\partial P_U}$. The optimal $P_U(H_U)$ to problem (P3)   is obtained as
\begin{small}
\begin{equation}\label{eq: suboptimal_P_U}
P_U(H_U)=\begin{cases}
P_{\textrm{max}}, &\textrm{if~}g(P_{\textrm{max}})> 0, \\
P_s, &\textrm{~otherwise},
\end{cases}
\end{equation}
\end{small}where $P_U\!=\!P_s$  is the unique solution to $ g(P_U)\!=\!0$ when $g(P_{\textrm{max}})\!\leq \!0$.
\end{proposition}

Proof to Proposition \ref{proposition: suboptimal_P_U} is  obtained   by   showing  that   $g(P_U)\!>\!0$   at $P_U\!=\!0$  and $g(P_U)$  monotonically decreases over  $P_U$  since $\frac{d g(P_U)}{d P_U}\!<\!0$,  which leads to the conclusion that    $\widetilde{\textrm{EE}}_U$   increases  over $P_U\!\in\![0, P_{\textrm{max}}]$ if $g(P_{\textrm{max}})\!>\!0$, or otherwise, $\widetilde{\textrm{EE}}_U$  first increases and then decreases over $P_U$,  and thus  (\ref{eq: suboptimal_P_U})  follows, while the details are omitted   for brevity.

By substituting (\ref{eq: suboptimal_P_U}) into (\ref{eq: EE_U_approx_complete}), the optimal $H_U^{EE}$ and thus $P_U^{EE}\!=\!P_U(H_U^{EE})$ to problem (P3) are then easily obtained via one-dimensional exhaustive search over  $H_U\in [H_0,H_{\textrm{max}}]$.  Hence,  given a quantization step size $\delta>0$, the complexity for solving (P3) is  only of  $\mathcal{O}(N)$, where   $N=\lfloor\frac{H_{\textrm{max}}-H_0}{\delta}\rfloor$   and  $\lfloor x \rfloor$ denotes  the floor operation of $x\in\mathbb{R}$.

Finally, as detailed in  Algorithm 1, by adopting  the optimal $H_U^{EE}$ and $P_U^{EE}$  to problem (P3)  as suboptimal solutions to problem   (P2), the original problem (P1) is   solved  with suboptimal $P_U^{EE}$, $H_U^{EE}$, $P_F^{EE}$,  and $\rho^{EE}$, where  $P_F^{EE}$ and $\rho^{EE}$ are     obtained by   substituting $P_U^{EE}$ and $H_U^{EE}$ into  (\ref{eq: optimal_FSO}).

\begin{algorithm}
\small
\caption{Proposed algorithm for  solving problem (P1).}
\begin{algorithmic}[1]
\small
\STATE initialize $\delta$  and $\textrm{EE}_{max}=0$.
\FOR{each $H_U\in\{H_0, H_0+\delta,..., H_0+N\delta\}$}
\STATE set $P_U(H_U)$ as in (\ref{eq: suboptimal_P_U})   and  $\widetilde{\textrm{EE}}_U$   as in (\ref{eq: EE_U_approx_complete});
\IF{$\widetilde{\textrm{EE}}_U>\textrm{EE}_{max}$}
\STATE set $\textrm{EE}_{max}=\widetilde{\textrm{EE}}_U$, $H_U^{EE}=H_U$, $P_U^{EE}=P_U(H_U)$;
\ENDIF
\ENDFOR
\STATE set $P_F^{EE} $ and $\rho^{EE}$ as in (\ref{eq: optimal_FSO}) using $P_U^{EE}$ and $H_U^{EE}$;
\STATE return  $P_U^{EE}$, $H_U^{EE}$, $P_F^{EE}$,  and $\rho^{EE}$.
\end{algorithmic}
\end{algorithm}

Thanks to the  closed-form expressions in (\ref{eq: optimal_FSO}),   the complexity of Algorithm 1 is also   of $\mathcal{O}(N)$ only, which equals  that for  solving problem (P3).   Therefore,  problem  (P1)  is efficiently  solved by our proposed Algorithm 1 with low complexity.

\section{Numerical Results}
This section provides numerical results   to validate our analysis and evaluate the proposed solution. We consider $\beta\!=\!15$ dB, $\kappa\!=\!4.3\!\times \!10^{-4}$/m, $\theta_{\textrm{t}}\!=\!0.06$ rads, $D_{\textrm{r}}\!=\!0.2$ m,  $\eta\!=\! 0.2$, and $\tau_{e}\tau_{GML}\!=\!0.9$ based on \cite{Schulz.LighWJ.2016} and \cite{FSO_book}. Unless specified otherwise, we  set $H_{\textrm{max}}\! =\! 200$, $L_0\! =\! 150$,   $R_0\! =\! 50$, all in meters,  $\alpha_{\textrm{L}}\! =\! 3$, $\alpha_{\textrm{N}}\!=\!5$, $\sigma_{\textrm{L}}^2\!=\! \sigma_U^2\! =\! 10^{-6}$ mW, $\sigma_{\textrm{N}}^2 \!= \!0.8 \!\times \! 10^{-6}$ mW, $\delta=1$,$P_{\textrm{max}}\!=\! 200$ mW, and $P_{\textrm{hov}}=1$ KW as in \cite{Che.TWC.2021}.

\begin{figure}[t]
    \centering
    \includegraphics[width=0.36\textwidth]{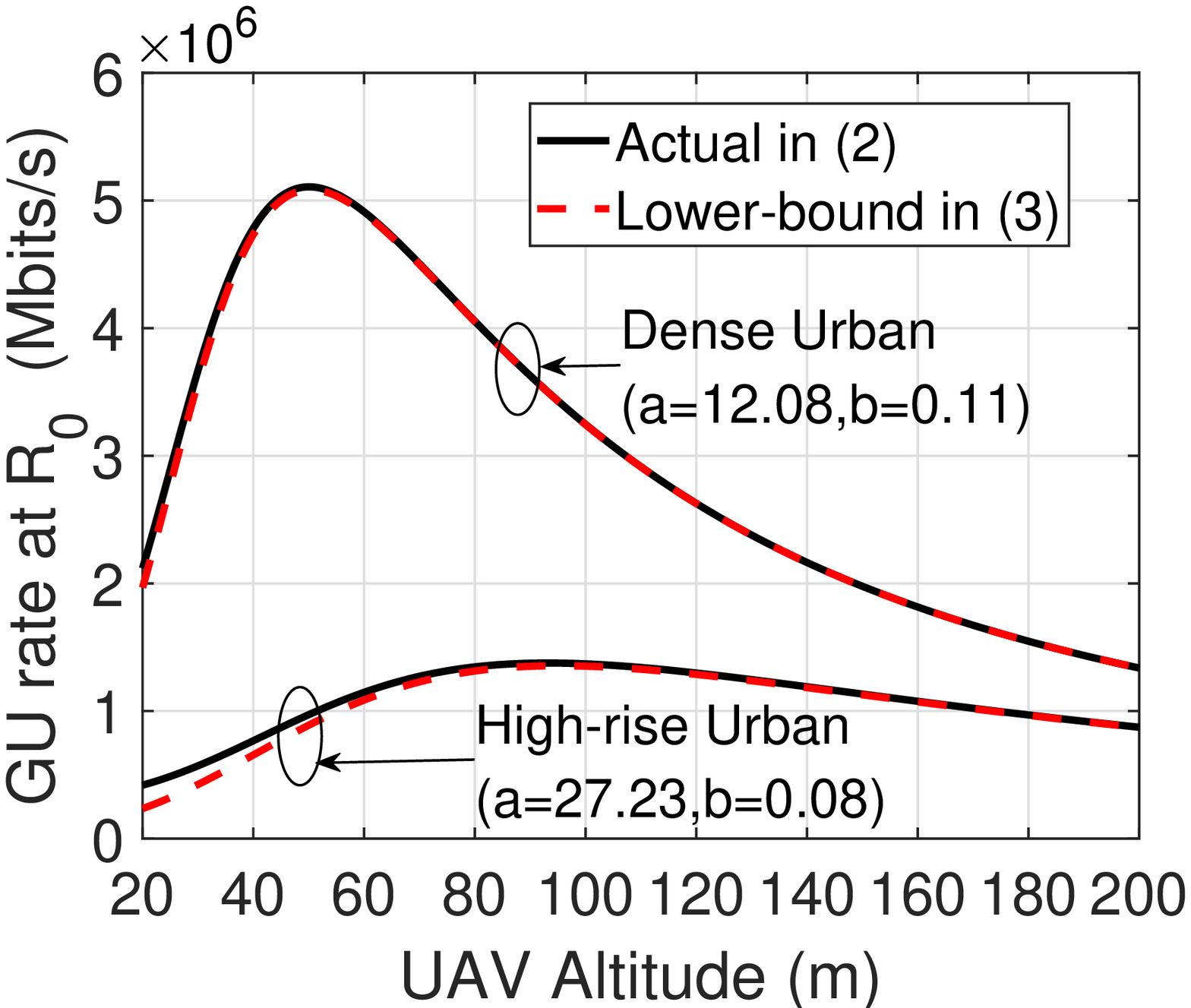}
     \vspace{-0.05in}
    \caption{ $\bar{C}(R_0)$ and its lower-bound $\bar{C}_{low}(R_0)$ over $H_U$.}
    \label{fig: Rate_approximate}
   \vspace{-0.05in}
\end{figure}

First, we validate that    $\bar{C}_{low}(r)$ in (\ref{eq: C_low}) is close to  $\bar{C}(r)$  in  (\ref{eq: C}) when $H_U$ is sufficiently large, under the typical environment  of  Dense Urban or High-rise Urban   in \cite{Hourani.letter.2014}  with generally low  LoS probabilities.
For each  environment, we conduct a worst-case study by focusing on  a GU   at radius $R_0$, which has  the lowest  LoS probability    and thus   the largest rate gap  $\bar{C}(r)\!-\!\bar{C}_{low}(r)$ among all $r\in[0,R_0]$ for any given $H_U$.
From  Fig.~\ref{fig: Rate_approximate}, it is first observed  that $\bar{C}(R_0)$ in the  Dense Urban case is always higher than that in the High-rise Urban case, due to the larger LoS probability in the former case. 
It is also observed that with $B=20$ MHz, the lower bound $\bar{C}_{low}(R_0)$ is always tight to the actual $\bar{C}(R_0)$ in the Dense Urban case.
 In the High-rise Urban case, although the rate gap   is  relatively larger  when $H_U$ is small, as the LoS probability increases over $H_U$,    the rate gap  becomes trivial when $H_U$ is sufficiently large. For example, when $H_U\!\geq\! 60$ m, the normalized rate gap with  $\frac{\bar{C}(R_0)\!-\!\bar{C}_{low}(R_0)}{\bar{C}(R_0)}\!\leq\! 0.052$ is obtained for the considered worst case.
For all  simulations in the following, we consider the High-rise Urban environment   and set  $H_0=60$ m.

\begin{figure*}[t]
\hspace{0.3cm}
\begin{minipage}[t]{0.3\linewidth}
    \centering
    \includegraphics[width=  1\textwidth]{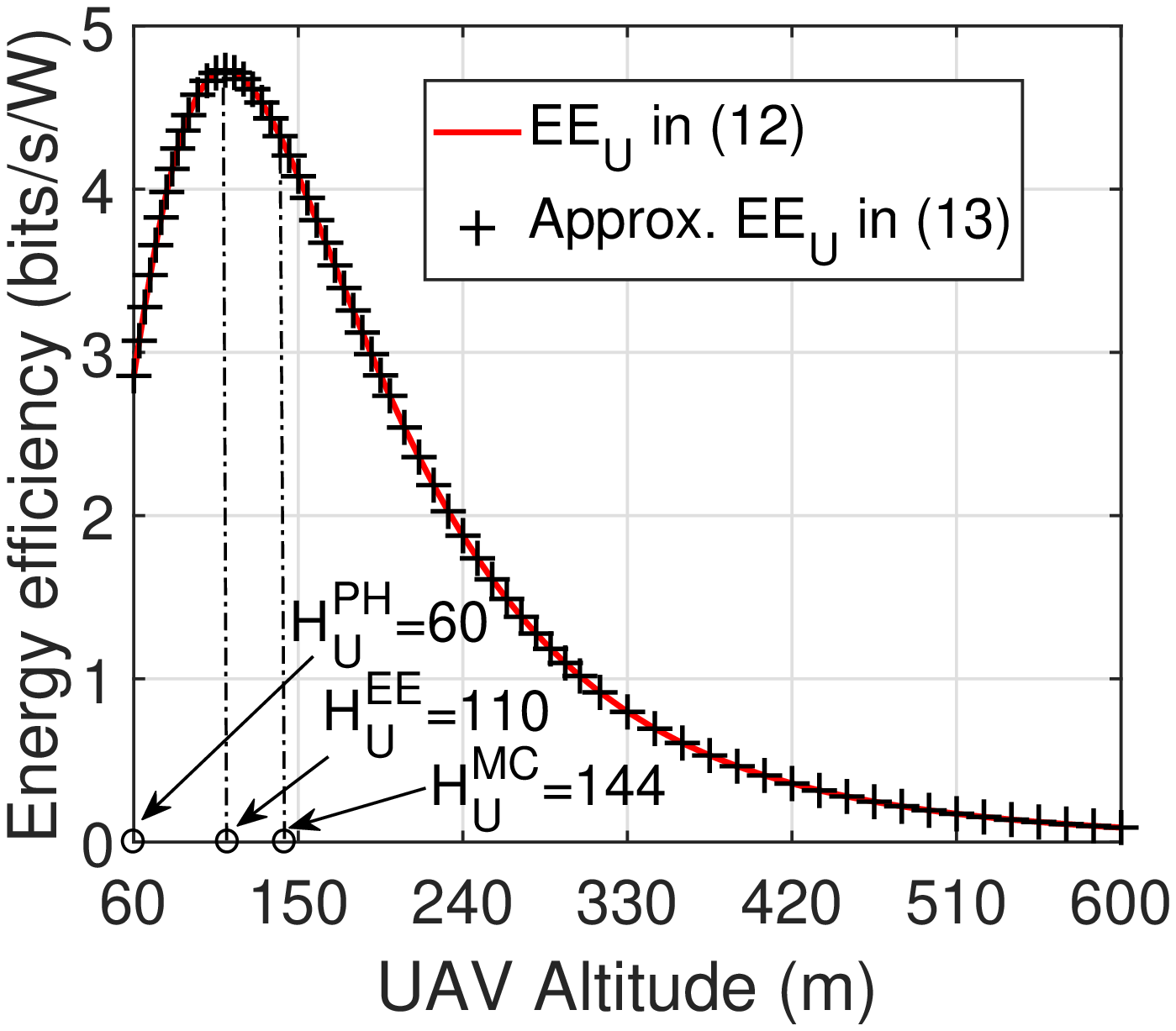}
    \vspace{-0.3in}
    \caption{  $\textrm{EE}_U$ and $\widetilde{\textrm{EE}}_U$ over $H_U$.}
    \label{fig: EE_over_H}
\end{minipage}
\hspace{0.1cm}
\begin{minipage}[t]{0.3\linewidth}
    \centering
    \includegraphics[width=  1\textwidth]{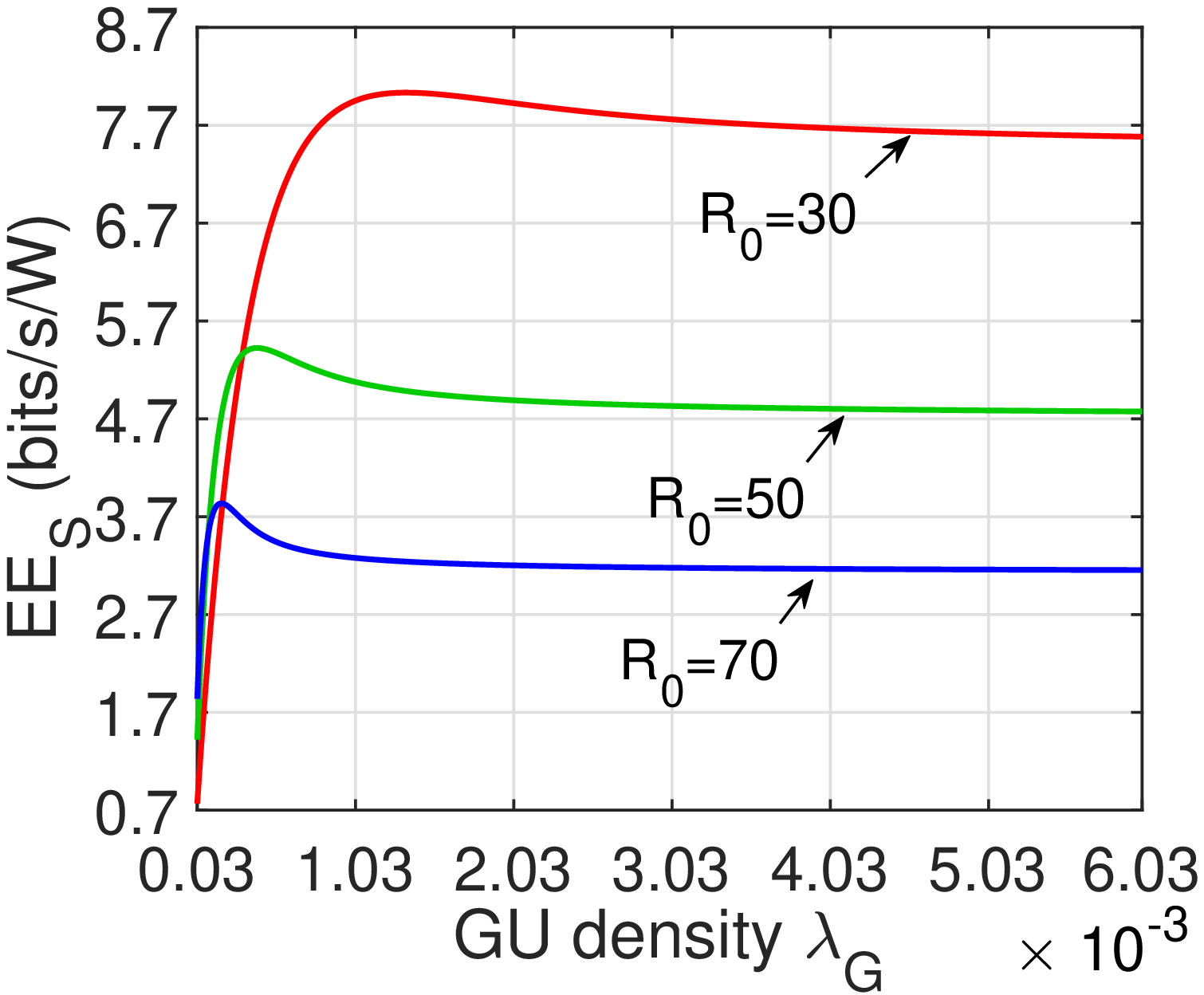}
    \vspace{-0.3in}
    \caption{System-level EE over $\lambda_G$.}
    \label{fig: EE_over_lambda_G}
\end{minipage}
\hspace{0.1cm}
\begin{minipage}[t]{0.3\linewidth}
    \centering
    \includegraphics[width= 1\textwidth]{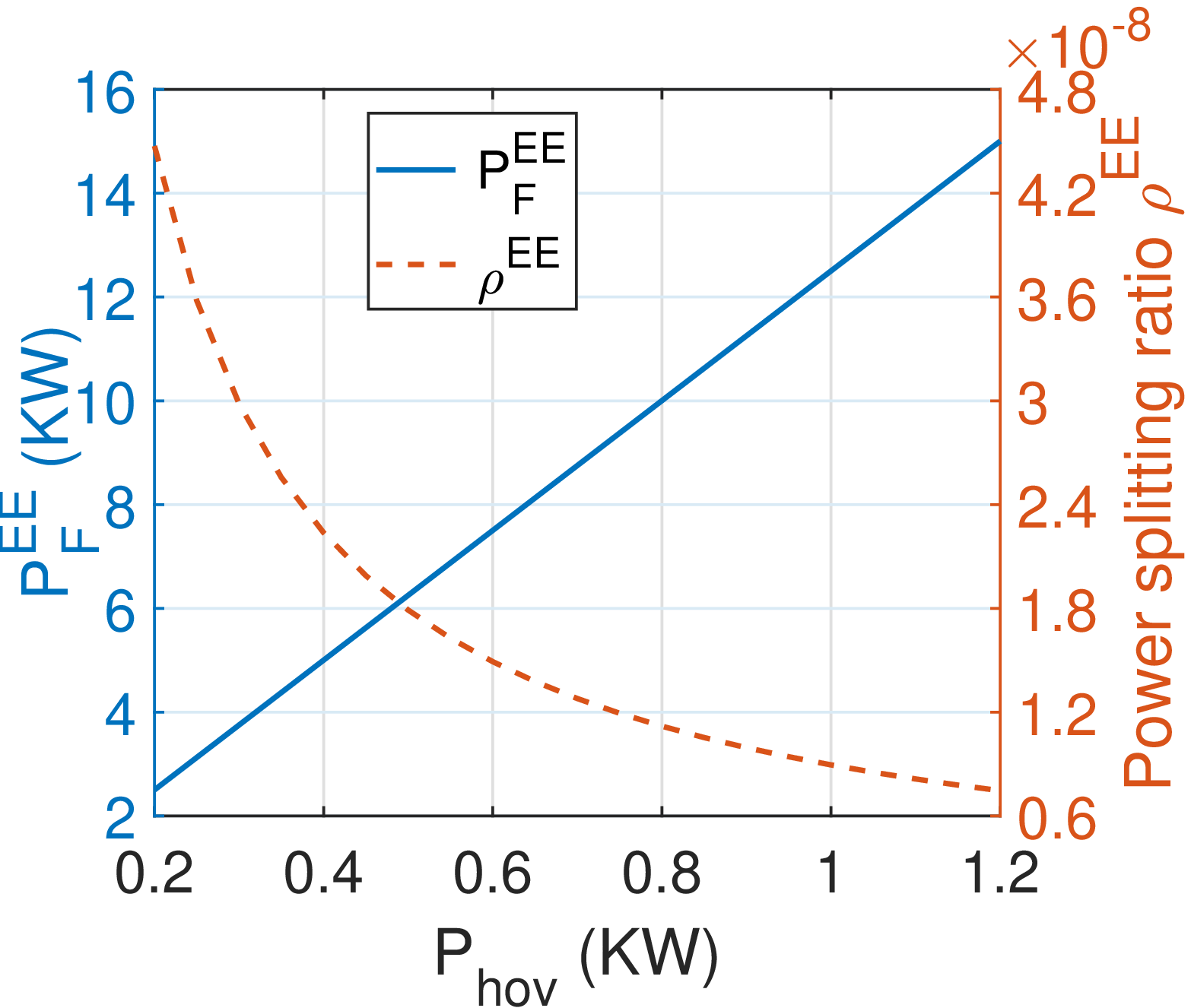}
    \vspace{-0.3in}
    \caption{$P_F^{EE}$ and $\rho^{EE}$ over $P_{\textrm{hov}}$.}
    \label{P_laser_and_rho_over_P_hover.eps}
\end{minipage}
\end{figure*}
 
 Next, we validate  that  $\widetilde{\textrm{EE}}_U$ in (\ref{eq: EE_U_approx})  is a good approximation to   $\textrm{EE}_U$ in (\ref{eq: EE_U}) in Fig.~\ref{fig: EE_over_H}. We set   $H_{\textrm{max}}\!=\!600$ m.
It is first found that at each value of $H_U$,  the optimal $P_U$  that maximizes $\widetilde{\textrm{EE}}_U$ with $\bar{C}_{edge}^{upp}$ or $\textrm{EE}_U$ with $\bar{C}_{edge}$, which is obtained  based on (\ref{eq: suboptimal_P_U}) or via one-dimensional search, is equal to $P_{\textrm{max}}$.
Under the optimal $P_{\textrm{max}}$, it is then observed that, as analyzed in Section III,
due to the dominance of   the   large $P_{\textrm{hov}}$,  $\widetilde{\textrm{EE}}_U$   is only trivially larger than  $\textrm{EE}_U$  at each $H_U$, where the resultant EE gap is  $\widetilde{\textrm{EE}}_U\!-\!\textrm{EE}_U \!=\! 9.1172\times10^{-7}$ at its largest value over all $H_U$'s.
It is further observed that  as  $H_U$ increases, both $\textrm{EE}_U$  and $\widetilde{\textrm{EE}}_U$  first  increase  due to the increased LoS probability, and then decrease after achieving their respective maximum   due to the significantly decreased $\omega(H_U)$.  In addition,  let  $H_U^{PH}$ and $H_U^{MC}$ denote the UAV's altitudes that maximize    $P_{rec}$ in (\ref{eq: P_rec}) and   $\bar{C}_{edge}$ in (\ref{eq: C_edge}), respectively. It is observed that $H_U^{PH} \!<\!H_U^{EE}\!<\! H_U^{MC}$. Therefore, the proposed   $H_U^{EE}$  for EE maximization  strikes a proper balance between FSO power and MC rate maximizations.

 Fig.~\ref{fig: EE_over_lambda_G} shows   $\textrm{EE}_S$  in (\ref{eq: EE})   over the GU density $\lambda_G$ by applying Algorithm 1.
 For each value of $R_0$,  it is observed that  $\textrm{EE}_S$    increases when $\lambda_G$, where   $\bar{C}_{edge}$ in (\ref{eq: C_edge}) increases fast, and after achieving its maximum value (at, e.g., $\lambda_G=1.23\times 10^{-3}$ when $R_0=30$), $\textrm{EE}_S$  begins to decrease   over $\lambda_G$,  due to  the resultant  increasing of   $R_{edge}$   and thus  $H_U^{EE}$  to enhance the LoS probability,   which in turn leads to   dominantly increased OBS transmit power  $P_F$ over  $\bar{C}_{edge}$  to  meet  (\ref{eq: LEH_constraint}). Moreover, when $\lambda_G$  is sufficiently large such that $R_{edge}$ equals $R_0$ with probability close to $1$, $H_U^{EE}$ does not increase any more and thus $\textrm{EE}_S$   becomes   a constant in Fig.~\ref{fig: EE_over_lambda_G}.
 Similarly,   when $\lambda_G$ is very small,  $\textrm{EE}_S$ is dominated  by the increasing of $\bar{C}_{edge}$ and thus increases over $R_0$; and when $\lambda_G$ is large,   $\textrm{EE}_S$ is  dominated by the increasing  of $P_F$  and thus generally decreases over $R_0$.
 In addition, it is also  observed from  Fig.~\ref{fig: EE_over_lambda_G} that  the optimal GU density that maximizes  $\textrm{EE}_S$ increases as $R_0$ decreases. This is   due to the increased   $\bar{C}_{edge}$ in (\ref{eq: C_edge}) under a smaller $R_0$,   which allows the UAV to serve more GUs at a lower altitude $H_U^{EE}$ using a smaller $P_F$   under (\ref{eq: backhaul_constraint}) and  (\ref{eq: LEH_constraint}).

 Finally, Fig.~\ref{P_laser_and_rho_over_P_hover.eps} shows $P_F^{EE}$ and $\rho^{EE}$ over  $P_{\textrm{hov}}$, where  $P_U^{EE}\!=\!P_{\textrm{max}}$ at each $P_{\textrm{hov}}$. Since $P_{\textrm{hov}}\!\gg \!P_U^{EE}$   in practice,  the UAV's  harvested FSO power is mainly used to satisfy  $P_{\textrm{hov}}$.
 This explains the generally low $\textrm{EE}_S$ in Figs. \ref{fig: EE_over_H} and \ref{fig: EE_over_lambda_G}, which further shows the importance
 of system-level EE design for FSO-powered  UAV communications.
Although   $\rho^{EE}$    is very small, due to the large $P_F^{EE}$,  the achieved SNR
 at the UAV is still acceptable with, e.g., $\frac{P_F \omega (H_U) \rho}{\sigma_U^2 \beta} \!=\!1.42$ in (\ref{eq: D_back}) when $P_{\textrm{hov}}=1$ KW.

\section{Conclusion}

This letter proposed a sustainable and reliable UAV MC system enabled by simultaneous FSO backhaul and power transfer.
We investigated the important system-level EE maximization design by jointly optimizing the FSO and RF transmissions and the UAV altitude. Despite of the non-convexity of the proposed problem, we obtained close-to-optimal solutions with low complexity  by tightly approximating the complicated EE. It was shown that the proposed design can properly balance the FSO power transfer and the RF downlink  MC.
It is our hope that this work can provide design insights on the FSO-based sustainable and reliable UAV RF access network.  It is also  an interesting future direction to extend this work to  mobile UAV scenarios by invoking  UAV trajectory  design to further improve the EE for FSO-enabled   UAV communications.

\appendices
\section{Proof to Proposition~\ref{proposition: C_upper_bound}}
When  $H_U\!>\!\sqrt{\alpha_{\textrm{L}}\!+\!1}R_0$, since $(H_U^2\!+\!r^2)^{\!-\frac{\alpha_{\textrm{L}}}{2}}$ is concave over $r$  with $\frac{\partial ^2}{\partial r^2} (H_U^2\!+\!r^2)^{\!-\frac{\alpha_{\textrm{L}}}{2}} \!<\!0$, and  the concave function $\log_2(1\!+\!\frac{P_U}{\sigma_{\textrm{L}}^2}X)$  monotonically increases over $X\!>\!0$,  we obtain that
   $Z(r)\!\triangleq\! \log_2[1\!+\!\frac{P_U}{\sigma_{\textrm{L}}^2}(H_U^2\!+\!r^2)^{-\!\frac{\alpha_{\textrm{L}}}{2}}]$ is concave over $r\!>\!0$.
Then, defining a  pdf  of $r\! \in \![0,R_0]$ as $f^{'}(r)\!=\!\frac{1}{\mu}\mathcal{P}_{L}(r)f_{R_{edge}}(r)$ with $\int_{0}^{R_0} \! f^{'}(r) \!=\!1$, we
obtain from  (\ref{eq: C_edge}) that $\bar{C}_{edge} \!= \!\mu B \int_0^{R_0} \! f^{'}(r)  Z(r)\,dr  \! \leq \! \bar{C_{edge}^{upp}}$, by applying Jensen's inequality with  the concave    $Z(r)$. Proposition~\ref{proposition: C_upper_bound} thus follows.

\end{document}